\documentclass[8.5pt,twoside,twocolumn]{article}
\usepackage[super,sort&compress,comma]{natbib} 
\usepackage{mhchem}
\usepackage{times,mathptmx}
\usepackage{sectsty}
\usepackage{balance} 
\usepackage{color}

\usepackage{graphicx} %eps figures can be used instead
\usepackage{epstopdf}
\usepackage{amsmath}
\usepackage{lastpage}
\usepackage[format=plain,justification=raggedright,singlelinecheck=false,font=small,labelfont=bf,labelsep=space]{caption} 
\usepackage{fancyhdr}
\makeatletter
\DeclareRobustCommand\onlinecite{\@onlinecite}
\def\@onlinecite#1{\begingroup\let\@cite\NAT@citenum\citealp{#1}\endgroup}
\makeatother   

\begin{document}

% Math Macros
\def\openone{\leavevmode\hbox{\small1\kern-3.3pt\normalsize1}}
\def\half{ \frac{1}{2}}
\def\eps{\varepsilon}
\def\vphi{\varphi}

\newcommand{\Op}[1]{\mathbf{\hat{#1}}}
\newcommand{\vOp}[1]{\mathbf{\hat{\vec{#1}}}}
\newcommand{\dotOp}[1]{\dot{\hat{\mathsf{\boldsymbol{#1}}}}}
\newcommand{\ddotOp}[1]{\ddot{\hat{\mathsf{\boldsymbol{#1}}}}}
\newcommand{\Opnohat}[1]{\boldsymbol{\mathsf{#1}}}
\newcommand{\Fkt}[1]{\,\mathsf {#1}}
\newcommand{\e}{\Fkt{e}}
\ifx\Tr\renewcommand{\Tr}{\Fkt{Tr}}
\else\newcommand{\Tr}{\Fkt{Tr}}
\fi
\newcommand{\diff}{\Fkt{d}}
\newcommand{\sub}[1]{_{\mathsf {#1}}}
\newcommand{\up}[1]{^{\mathsf {#1}}}
\renewcommand{\dagger}{+}
\newcommand\hateq{\ensuremath{\hat =}}

\thispagestyle{plain}
\fancypagestyle{plain}{
%\fancyhead[L]{\includegraphics[height=8pt]{headers/LH}}
%\fancyhead[C]{\hspace{-1cm}\includegraphics[height=20pt]{headers/CH}}
%CK\fancyhead[R]{\includegraphics[height=10pt]{headers/RH}\vspace{-0.2cm}}
\renewcommand{\headrulewidth}{1pt}}
\renewcommand{\thefootnote}{\fnsymbol{footnote}}
\renewcommand\footnoterule{\vspace*{1pt}% 
\hrule width 3.4in height 0.4pt \vspace*{5pt}} 
\setcounter{secnumdepth}{5}

\makeatletter 
\def\subsubsection{\@startsection{subsubsection}{3}{10pt}{-1.25ex plus -1ex minus -.1ex}{0ex plus 0ex}{\normalsize\bf}} 
\def\paragraph{\@startsection{paragraph}{4}{10pt}{-1.25ex plus -1ex minus -.1ex}{0ex plus 0ex}{\normalsize\textit}} 
\renewcommand\@biblabel[1]{#1}            
\renewcommand\@makefntext[1]% 
{\noindent\makebox[0pt][r]{\@thefnmark\,}#1}
\makeatother 
\renewcommand{\figurename}{\small{Fig.}~}
\sectionfont{\large}
\subsectionfont{\normalsize} 

\fancyfoot{}
%\fancyfoot[LO,RE]{\vspace{-7pt}\includegraphics[height=9pt]{headers/LF}}
%\fancyfoot[CO]{\vspace{-7.2pt}\hspace{12.2cm}\includegraphics{headers/RF}}
%\fancyfoot[CE]{\vspace{-7.5pt}\hspace{-13.5cm}\includegraphics{headers/RF}}
\fancyfoot[RO]{\footnotesize{\sffamily{1--\pageref{LastPage} ~\textbar  \hspace{2pt}\thepage}}}
\fancyfoot[LE]{\footnotesize{\sffamily{\thepage~\textbar\hspace{3.45cm} 1--\pageref{LastPage}}}}
\fancyhead{}
\renewcommand{\headrulewidth}{1pt} 
\renewcommand{\footrulewidth}{1pt}
\setlength{\arrayrulewidth}{1pt}
\setlength{\columnsep}{6.5mm}
\setlength\bibsep{1pt}
%%%%%%%%%%%%%%%%%%%%%%%%%%%%%%%%%%%%%Title%%%%%%%%%%%%%%%%%%%%%%%%%%%%%%%%%%%%%%%%%%%%%%%%
\twocolumn[
  \begin{@twocolumnfalse}
\noindent\LARGE{\textbf{Controlling a diatomic shape resonance with
    non-resonant light}}
\vspace{0.6cm}

%%%%%%%%%%%%%%%%%%%%%%%%%%%%%%%%%%%%%%Autors%%%%%%%%%%%%%%%%%%%%%%%%%%%%%%%%%%%%%%%%
\noindent\large{\textbf{Ruzin A\u{g}ano\u{g}lu,\textit{$^{a}$} 
  Mikhail Lemeshko,\textit{$^{b}$} Bretislav Friedrich,\textit{$^{b}$}
  Rosario Gonz\'alez-F\'erez,\textit{$^{c}$}
  Christiane P. Koch\textit{$^{a,d}$}}}\vspace{0.5cm} 

%CK\noindent\textit{\small{\textbf{Received Xth XXXXXXXXXX 2011, Accepted
  %CK    Xth XXXXXXXXX 2011\newline 
    %CK  First published on the web Xth XXXXXXXXXX 2011}}}

%CK \noindent \textbf{\small{DOI: 10.1039/b000000x}}
\vspace{0.6cm}

%%%%%%%%%%%%%%%%%%%%%%%%%%%%%%%%%%%Abstract%%%%%%%%%%%%%%%%%%%%%%%%%%%%%%%%%%%%%%%%%%%%%
\noindent \normalsize{
  A (diatomic) shape resonance is a metastable state of a pair of
  colliding atoms quasi-bound by the centrifugal barrier imposed by the
  angular momentum involved in the collision. The temporary trapping of
  the atoms' scattering wavefunction corresponds to an enhanced
  atom pair density at low interatomic separations. This leads to larger
  overlap of the wavefunctions involved in a molecule formation
  process such as photoassociation, rendering
  the process more efficient. However, for
  an ensemble of atoms, the atom pair density will only be enhanced if
  the energy of the resonance comes close to the temperature of the
  atomic ensemble. Herein we explore the possibility of controlling the energy of
  a shape resonance by shifting it toward the temperature of
  atoms confined in a trap. The shifts are imparted by the interaction
  of non-resonant light with the anisotropic polarizability of the atom
  pair, which affects both the centrifugal barrier and the pair's
  rotational and vibrational levels. 
  We find that at laser intensities of up to $5\times 10^{9}$ W/cm$^2$
  the pair density is increased by one order of magnitude 
  for $^{87}$Rb atoms at  $100~\mu$K and by two orders of magnitude
  for $^{88}$Sr atoms at  $20~\mu$K. 
}
\vspace{0.5cm}
\end{@twocolumnfalse}
]

\footnotetext{\textit{$^{a}$~Institut f\"ur Theoretische Physik, Freie
    Universit\"at Berlin, Arnimallee 14, 14195 Berlin, Germany.}}
\footnotetext{\textit{$^{b}$~Fritz-Haber-Institut der
    Max-Planck-Gesellschaft, Faradayweg 4-6, 14195 Berlin, Germany.}}
\footnotetext{\textit{$^{c}$~Instituto ‘Carlos I’ de F\'isica
    Te\'orica y Computacional and Departamento de F\'isica At\'omica,
    Molecular y Nuclear, Universidad de Granada, 18071 Granada,
    Spain.}} 
\footnotetext{\textit{$^{d}$~Theoretische Physik, Universit\"at
    Kassel, Heinrich-Plett-Str. 40, 34132 Kassel, Germany.
    Email: christiane.koch@uni-kassel.de
  }} 

%%%%%%%%%%%%%%%%%%%%%%%%%%%%%%%%%%Main%%%%%%%%%%%%%%%%%%%%%%%%%%%%%%%%%%%%%%%%%%%%%%%%%%

\section{Introduction}
\label{sec:intro}

The quest for translationally ultracold ($T\le 100\,\mu$K) molecules
in their absolute internal ground state has been rewarded recently
when ultracold atoms were associated via  a 
magnetic-field controlled Feshbach resonance followed by Stimulated 
Raman Adiabatic Passage.~\cite{NiSci08,DanzlSci08} 
Ultracold molecules were also produced by associating atoms using laser
light~\cite{JonesRMP06} and brought into their rovibrational ground
state by spontaneous or stimulated
emission.~\cite{SagePRL05,DeiglmayrPRL08,ViteauSci08}  
Feshbach or magneto-association requires non-zero nuclear spin -- 
a hyperfine manifold -- of the colliding
atoms.~\cite{KoehlerRMP06} It 
is efficient when the collision energy of the atoms is of the 
order of the hyperfine splitting. This corresponds to nano-Kelvin
temperatures which are reached via evaporative or sympathetic
cooling, leaving typically only about $10^{4}-10^{5}$ atoms that can be
associated. 
Photoassociation~\cite{FrancoiseReview,JonesRMP06} relies only on the presence
of optical transitions which are usually abundant, and is not tied to
any particular temperature regime (although  the
specific photoassociation mechanisms may differ at high~\cite{MarvetCPL95}
and very low~\cite{FrancoiseReview} temperatures).
In the ultracold domain, photoassociation has been often implemented in
combination with magneto-optical traps (MOTs),~\cite{FrancoiseReview}
which hold up to $10^{10}$ atoms at temperatures
ranging between $10-100\,\mu$K. However, only a small fraction of the
MOT atoms are at  internuclear separations, $R$, amenable to
photoassociation, i.e., the density of atom pairs within the required
range of $R$ is tiny. 
This reflects the fact that photoassociation cannot achieve the phase
space compression required to transform far-apart atoms into tightly
or even weakly bound molecules. 
%% The phase space argument is essential since it is not appreciated
%% by MOST people having published on photoassociation recently
%% (Yelin, Shapiro, ...)
As a result,  the number of ground state molecules
produced via photoassociation has been limited to only about
$10^4$.~\cite{ViteauSci08,DeiglmayrPRL08} 

A remedy for the limited efficiency of photoassociation could be found 
within the field of coherent control, which employs quantum 
interference to constructively enhance a desired outcome of a
process while destructively suppressing all its undesirable
alternatives.~\cite{RiceBook,ShapiroBook} 
Although coherent control has
been successful for unimolecular processes such as
photoionization or photodissociation,~\cite{RabitzSci00,Sfb450book}
controlling a binary reaction has remained an open challenge. This has
been mainly due to the fact that the initial state of the reaction
consists of an incoherent mixture of scattering states with random relative phases.
However, resonances can, in principle, endow the mixture of the
initial scattering states with a single phase.~\cite{ZemanPRL04} 

The way toward solving the problem of limited photoassociation
efficiency may be paved by considering photoassociation as a coherent 
control problem.
Coherent control employs quantum 
interference to constructively enhance a desired outcome of a
process while destructively suppressing all its undesirable
alternatives.~\cite{RiceBook,ShapiroBook}  
Although coherent control has
been highly successful for unimolecular processes such as
photoionization or photodissociation,~\cite{RabitzSci00,Sfb450book} 
controlling a binary reaction such as photoassociation remains an open challenge. 
%% It still is an open challenge!
This is mainly due to the fact that the initial state of the reaction
consists of an incoherent mixture of scattering  
states with random relative phases.
However, resonances can, in principle, endow the mixture of the
initial scattering states with a single phase.~\cite{ZemanPRL04} 

%%At ultralow temperatures the role of resonances is amplified by the
% Resonances help at high temperature but are even more important at
% low temperature, this needs to be said
%%fact that degrees of freedom are successively frozen out as the energy
%%scales become smaller and smaller. Eventually, the dynamics is 
%%governed by quantum effects such as resonances and tunneling. The
The importance of resonances increases in the ultracold regime where,
along with tunneling, they dominate the quantum dynamics. 
The presence of a Feshbach resonance has been  predicted to significantly enhance
the photoassociation yield in Feshbach optimized
photoassociation (FOPA).~\cite{PellegriniPRL08} Similarly, 
an enhanced microwave absorption is expected near a Feshbach
resonance.~\cite{AlyabyshevPRA10} However, Feshbach enhancement of
photoassociation is restricted to atoms with non-zero nuclear spin and
is the more efficient the lower the temperature. Although electric field-induced
resonances~\cite{KremsPRL06,RosarioNJP09,ChakrabortyJPB11} should also 
icnrease the photoassociation efficiency, the electric fields
required to achieve a significant enhancement are currently experimentally unfeasible.
Shape resonances that occur when a scattering state becomes
trapped behind the centrifugal barrier for partial waves with $J>0$
were found to yield enhanced photoassociation
rates.~\cite{BoestenPRL96,BoestenPRA97,LondonoPRA10}  However, due
to the rotational excitation involved in generating the barrier, the
lowest energies at which shape 
resonances occur correspond typically to temperatures of a 
few milli-Kelvin. Therefore, the thermal weight of a shape resonance in a
much colder MOT or optical trap is quite small. 

Herein, we study the possibility of enhancing the thermal weight of a
scattering resonance by shifting its position (with respect to the
trap energy) by applying a non-resonant radiative field. The
interaction of such a field with the anisotropic polarizability of the
atom pair is of a universal character, independent of 
any particular energy level structure, frequency of the light (as long
as it is non-resonant), or the presence of a permanent dipole moment.  
Non-resonant or far-off-resonant light was used to manipulate molecular
alignment~\cite{ FriedrichPRL95,FriedrichJPC95, StapSeiRMP03} and has been
predicted to shift rotational and vibrational levels
~\cite{LemeshkoJPCA10} or to modify  
intermolecular interactions~\cite{Lemeshko11}. It has also been predicted to
cause peculiar effects  in atomic Bose-Einstein condensates such as
gravitational self-binding and supersolid-like
structures.\cite{OdellPRL00,GiovanazziPRL02,OdellPRL03}
The nonresonant polarizability interaction  creates an
effective rotational barrier for weakly bound molecules; and the  
subsequent `shaking' of the molecules imparted by the nonresonant
light could be employed to recover the 
vibrational probability density distribution of the corresponding bound atom
pair.\cite{FriedrichCCCC98,LemeshkoPRL09} 
The intuitive picture of an effective centrifugal barrier 
provided an impetus for the present study: if the non-resonant
light creates (for $J=0$) or modifies (for $J>0$) the centrifugal
barrier, it will also affect 
the position of a shape resonance. Application of a non-resonsant
field should therefore allow to control a shape
resonance. The core of our present study extends the previous work on
shifting weakly bound diatomic levels to shifting of quasi-bound
scattering states. 

The ubiquitous diatomic low-$J$ shape resonances are straightforward
to predict from the basic scattering properties of 
ultracold atoms.~\cite{GaoPRA98,GaoPRA09,LondonoPRA10} 
Specifically, we consider a $d$-wave shape resonance in
87-rubidium~\cite{BoestenPRA97} and  a $g$-wave shape resonance in
88-strontium. Rubidium is the drosophila of ultracold physics, whose
potential energy curves 
and polarizability anisotropy are accurately known. 
$^{88}$Sr$_2$ features in proposals
to test the time dependence of the electron-to-proton mass
ratio.~\cite{ZelevinskyPRL08} With no Feshbach resonances present, 
$^{88}$Sr$_2$ has been produced by photoassociation.~\cite{NagelPRL05,ZelevinskyPRL06}
We compare the prospects for controlling the shape resonances of rubidium
and strontium with non-resonant light and draw general conclusions about the
applicability and efficiency of this control scheme for producing
diatomics via photoassociation.

\section{Theoretical framework}
\label{sec:theo}

\subsection{Description of a thermal cloud of atoms}
\label{subsec:rho}

The initial state of an ensemble of atoms held at thermal equilibrium in a
trap of temperature $T$ is described  by the
canonical density operator, 
\begin{equation}
  \label{eq:rho}
  \Op \rho_{T,N} = \frac{1}{Z}e^{-\Op H_N/k_B T}\,.
\end{equation}
Equation~\eqref{eq:rho} disregards the effects of quantum statistics,
which are negligible at typical MOT temperatures. 
In a dilute gas where three-body and higher order interactions can be
neglected, the Hamiltonian, $\Op H_N$, consists only of
single-particle ($\Op T$, $V_\mathsf{trap}$) and two-body $V_{ij}(\Op R_{ij})$
operators. The $N$-particle density operator, $\Op\rho_{T,N}$, is then
given in terms of $N^2$-times the pair density operator,
$\Op\rho_{T,2}$.~\cite{KochJPhysB06}  
Assuming that the center of mass and internuclear degrees of freedom can be separated -- as is the case, e.g., in  a harmonic trap -- the center
of mass motion can be integrated analytically. 
By representing the Hamiltonian for the internuclear motion on a
finite-size coordinate-space grid with variables $(R,\theta,\phi)$,
and making use of the azimuthal symmetry, 
the density matrix of the initial state can be 
constructed in terms of the eigenfunctions, $\varphi_{nJ}(R,\theta)$,
of the pair Hamiltonian, $\Op H_2$, with eigenvalues $E_{nJ}$,~\cite{KochJPhysB06}  
\begin{equation}
  \label{eq:rho_grid}
  \rho_{T,2} (R,\theta) = \frac{1}{4\pi R^2} 
  \frac{\sum_{nJ}(2J+1)e^{-E_{nJ}/k_BT}|\varphi_{nJ}(R,\theta)|^2}
  {\sum_{nJ}(2J+1)e^{-E_{nJ}/k_B T}}\,.
\end{equation}
The grid needs to be sufficiently large in $R$ to approximate the
scattering continuum well.~\cite{ElianePRA04} 
A small number of partial waves is sufficient to ensure convergence of
the short-range part of $\rho_{T,2} (R,\theta)$,
since only a few partial waves are thermally populated at
very low temperatures.~\cite{KochJPhysB06} 
%The long-range part of $\rho_{T,2} (R,\theta)$ requires of the
%order of 150 partial waves for convergence for 

Formally, the time evolution of the cloud of atoms is determined by the
dynamics of the pair density operator, 
\begin{equation}
  \label{eq:rho_t}
  \Op \rho_{T,2}(t) = e^{-\frac{i}{\hbar}\Op H_2 t}  \Op \rho_{T,2}(t=0)
  e^{\frac{i}{\hbar}\Op H_2 t}\,,
\end{equation}
under the assumption that on the timescale of $t$, the time evolution is unitary, i.e., no dissipative mechanisms are present.
Any thermal expectation value is obtained as 
$\langle \Op A(t)\rangle=\Tr[\Op A \Op\rho_{T,2}(t)]$. For example,
the photoassociation 
probability is given by $\langle \Op P_e(t_f)\rangle$, where $\Op P_e$
denotes the projector onto the electronic state $e$ which is populated
by photoassociation and $t_f$ is some final time. 
The unitary time evolution in Eq.~\eqref{eq:rho_t} implies that it is not
necessary to solve the Liouville-von Neumann equation for the density
operator explicitly. Rather, a separate
propagation for each eigenstate in Eq.~\eqref{eq:rho_grid} is
sufficient to calculate thermal expectation values.~\cite{KochJPhysB06}
In particular, if only the short range part of the initial thermal
density is probed, only few partial waves suffice for
convergence. 
This is numerically much more efficient than solving the Liouville-von
Neumann equation since the number of terms in the sum of
Eq.~\eqref{eq:rho_grid}  is relatively small due to the narrow thermal
width of an initial state at low temperature. 

\subsection{Interaction of a diatom with non-resonant laser light}\label{subsec:model}

The Hamiltonian of an atom pair in its electronic ground state 
in the presence of a non-resonant laser 
field can be written as
\begin{equation}
\label{eq:2D_Hamil}
\Op H^I_2=\Op T_R+\frac{\Op{J}^2}{2\mu \Op R^2}+V_g(\Op R)
-\frac{2\pi I}{c}\left(\Delta\alpha(\Op
  R)\cos^2\Op\theta+\alpha_\perp(\Op R)\right)\,,
\end{equation}
where the first and second terms denote the vibrational and rotational
kinetic energies, respectively, and  
$V_g(\Op R)$ the field-free ground electronic 
potential energy curve. The last term of
Eq.~\eqref{eq:2D_Hamil} 
represents the interaction with a non-resonant laser field, with 
$I$ the laser intensity and  $\Delta\alpha(\Op R)=\alpha_\parallel(\Op
R)-\alpha_\perp(\Op R)$ the polarizability anisotropy, given
in terms of the perpendicular and parallel molecular polarizability components,
$\alpha_\perp(\Op R)$ and $\alpha_\parallel(\Op R)$. For homonuclear dimers,
the behaviour at large internuclear distances is given by 
Silberstein's expansion,~\cite{Silberstein1,Silberstein2,Silberstein3,JensenJCP01}
\begin{eqnarray}
 \alpha_\perp(\Op R) &\approx& 2\alpha_0-2\alpha_0^2/\Op
 R^3+2\alpha_0^3/\Op R^6\,, \nonumber \\
 \alpha_\parallel(\Op R) &\approx& 2\alpha_0+4\alpha_0^2/\Op R^3
 +8\alpha_0^3/\Op R^6\,,
 \label{eq:alpha}
\end{eqnarray}
%$\Delta\alpha(r)\approx 6\alpha_0^2/\Op R^3 +6\alpha_0^3/\Op R^6$
where $\alpha_0$ is the atomic polarizability.
Hamiltonian~\eqref{eq:2D_Hamil} is derived by assuming the
frequency of the laser to be far from any resonance and larger 
than the inverse of both the pulse duration and the rotational
period. In this case, a two-photon rotating-wave approximation averaging
over the rapid oscillations of the non-resonant field can be
applied.~\cite{PershanPR66} Furthermore, far from resonances, the
frequency-dependent molecular polarizability approaches its static
value, which allows to cast Eq.~(\ref{eq:2D_Hamil}) in the
static polarizability limit. The second-order nature of the 
light-matter interaction is reflected by the intensity (not the field
amplitude) and  $\cos^2 \Op{\theta}$ operator occurring in the
last term of Eq.~\eqref{eq:2D_Hamil}. 
Thus the energy of the field-dressed eigenstates always decreases with
increasing field intensity, i.e., the states are high-field seeking.

Since an external field defines a preferred direction in space, the
symmetry of the corresponding Hamiltonian is 
reduced compared to the field-free case. 
In the absence of a field, a bound molecular state is characterized by its
vibrational, rotational, and magnetic quantum numbers $(\nu, J,M)$.
This carries over to unbound box-discretized  states approximating the
scattering continuum above the dissociation limit, i.e., $(n, J,M)$
where each translational ``quantum number'' represents a range of
scattering states with collisional energies close to
$E_n$.~\cite{KochJPhysB06}
Due to the azimuthal symmetry about the laser polarization axis,
the light-matter interaction depends only on the polar angle
$\theta$. As a consequence, 
by turning on the non-resonant laser field, hybridization of the
rotational motion takes place and  only the magnetic quantum number
$M$ remains conserved.  
The field-free degeneracy of states with the same rotational
quantum number $J$ but different $M$ is lifted  
due to the interaction with the laser field. 
Herein, we focus on states with $M=0$, since for a given $J$  
the effect of the non-resonant field increases with decreasing $M$ and
is largest for $M=0$. 
Thus, the sum in the density matrix~\eqref{eq:rho_grid} describing the
initial state is restricted to 
states with $M=0$, setting the degeneracy factor $(2J+1)$ to 1.

Discrete basis set methods are employed to represent the vibrational
and rotational degrees of freedom of Hamiltonian
\eqref{eq:2D_Hamil}. For the radial part, a mapped Fourier grid is
used with the grid step set to be proportional to the local de Broglie
wavelength.~\cite{SlavaJCP99,WillnerJCP04,ShimshonCPL06}
This allows us to use large enough discretization boxes to
properly describe the part of the scattering continuum that is
relevant at ultracold temperatures.
The rotational  degree of freedom is treated by a discrete variable
representation in terms of Legendre polynomials, 
taking into account that $M$ is conserved.~\cite{bacic1989,light2000}
At large intensities, more partial waves than
those that are thermally populated initially will come into play,
cf. Eq.~\eqref{eq:rho_grid}. Convergence of our
calculations with respect to the number of
Legendre polynomials is ensured for the largest intensity
employed ($J^I_{max} \approx 14$ compared with $J^0_{max}=3$ or 4 in the
field-free case). 

\begin{figure}[tb]
  \centering
  \includegraphics[width=0.9\linewidth]{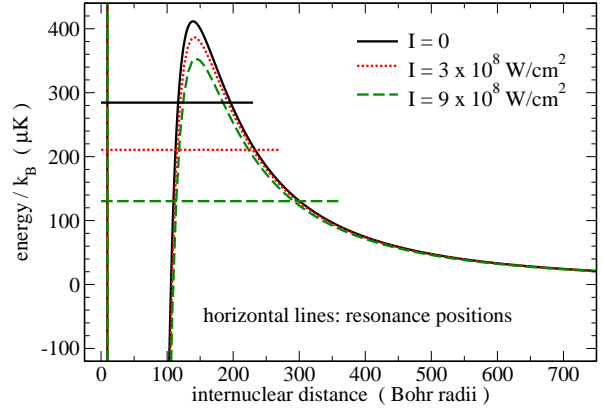}
  \caption{Effective potential, $V_g(R)+E_j(R;I)$, that the atom pair
    experiences in the 
    presence of non-resonant laser light. The field-free rotational
    quantum number is $J=2$ for which a shape resonance is observed at
    about $290\,\mu$K. The non-resonant light shifts the position of the
    shape resonance to lower
    energies as indicated by the vertical lines (resonance positions
    calculated with the full $2D$ Hamiltonian).}
\label{fig:eff_pot}
\end{figure}

As evident from Eq.~\eqref{eq:2D_Hamil}, the interaction with the
non-resonant laser light couples $\Op R$ and $\Op \theta$, i.e., it
affects both vibrational and rotational motion of the atom pair. In
order to ease interpretation, % of the effects of non-resonant light on the atom pair,
it is expedient to disentangle the two effects by making use of effective
one-dimensional models.
The effect on the rotational motion %after the non-resonant laser
                                %pulse is over 
is captured by an effective quantum number, $J^*$, determining
the rotational barrier in an effective one-dimensional (vibrational)
model.~\cite{LemeshkoPRL09} 
% \textbf{THIS IS ONLY IN THE NONADIABATIC CASE. In the adiabatic case
% $J^\ast \equiv 0$ after the pulse is gone (Misha)}
% CK : is this important here? 
If the non-resonant laser field is kept on, 
the effect of the field on the angular motion 
at a given internuclear separation needs to be accounted for by
including the admixture of different partial waves  and of the alignment. 
We therefore generalize previous
treatments~\cite{FriedrichPRL95,FriedrichJPC95,PhysRevA.69.023402,PhysRevA.71.033416,LemeshkoPRL09,LemeshkoJPCA10}  
to the latter case.
In order to obtain an effective one-dimensional model in the presence
of the non-resonant light, we first solve the angular part of the
eigenvalue problem for a fixed $R$, 
\begin{eqnarray}
  \label{eq:1D_Hamil_angular}
  \left[\frac{\Op{J}^2}{2\mu R^2}-\frac{2\pi I}{c}
    \left(\Delta\alpha(R)\cos^2\Op\theta+\alpha_\perp(R)\right)\right]
  \Phi_j(\theta;R,I) \nonumber \\ = E_j(R,I)\Phi_j(\theta;R,I)\,.
\end{eqnarray}
The eigenfunctions, $\Phi_j(\theta;R,I)$,
and eigenvalues, $E_j(R,I)$, of this equation depend
parametrically on the radial coordinate $R$ and on the  
laser intensity $I$. 
The index $j$ is a label which is related to the field-free rotational
quantum number. 
The full wavefunction can be expanded in terms of these angular
wavefunctions, 
\begin{equation}
  \label{eq:2D_wf_approx}
  \Psi_{nJ}(R,\theta;I)=\sum_{j'}\psi_{nj'}(R;I)\Phi_{j'}(\theta;R,I)\,.
\end{equation}
%Using this expansion in the Schr\"odinger equation associated to the
%Hamiltonian  \eqref{eq:2D_Hamil}, multiplying by  
%$\Phi_{j}^*(\theta;R)$ and integrating over the angular coordinate yields
%\begin{eqnarray}
%  \label{eq:2D_equa_r_pre}
%  \left[T_R+V(R)+E_j(R;I)-E\right]\psi_{j}(R;I) =  \nonumber \\
%  -\sum_{j'}\psi_{j'}(R,I)\int_0^\pi\Phi_{j}^*(\theta;R,I)
%  \Op T_R\Phi_{j'}(\theta;R,I)\sin\theta d\theta\,.
%\end{eqnarray}
%Neglecting all non-adiabatic couplings on the right hand side of
%Eq.~\eqref{eq:2D_equa_r_pre}, an effective one-dimensional
%Schr\"odinger equation for the vibrational part is 
%obtained,
Inserting this expansion into the time-independent Schr\"odinger
equation for Hamiltonian  \eqref{eq:2D_Hamil},
multiplying by $\Phi_{j}^*(\theta;R)$, integrating over the
angular coordinate $\theta$,
and neglecting all non-adiabatic coupling matrix elements,  
$\int_0^\pi\Phi_{j}^*(\theta;R,I)
\Op T_R\Phi_{j'}(\theta;R,I)\sin\theta d\theta$,
the following effective one-dimensional
Schr\"odinger equation for the vibrational part is obtained,
\begin{equation}
  \label{eq:1D_eff}
  \left[\Op T_R+V_g(\Op R)+E_j(\Op R;I)\right]\psi_{nj}(R;I) = E_n
  \psi_{nj}(R;I) \,.
\end{equation}
Here, $V_g(R)+E_j(R;I)$ is the effective potential
including the effect of non-resonant
light. Note that although 
the index $n$ is used as a label indicating a scattering state, 
this adiabatic model is also applicable to bound states. 
Figure~\ref{fig:eff_pot} compares the 
field-free potential, $V_g(R)+J(J+1)/(2\mu R^2)$ for $J=2$, with the 
effective potentials,  $V_g(R)+E_j(R;I)$, for two
intensities of the non-resonant field for $^{87}$Rb$_2$: as the
field intensity increases, the height of the rotational barrier
decreases. 

Equation~\eqref{eq:1D_eff} represents an adiabatic approximation
to the coupled rovibrational motion where the rotational 
wavefunction depends parametrically on the radial variable.
This approximation is valid if the energy scales, or, respectively,
timescales, associated with the rotational and vibrational
motion are well separated. For very large field intensities, the
adiabatic approximation is expected to break down, 
because the matter-field interaction becomes comparable to the
vibrational  energy scale. This allows for 
coupling between states in different 
vibrational bands which is neglected in Eq.~\eqref{eq:1D_eff}. 

\subsection{Envisaged scheme for enhancement of photoassociation}
%  by controlling a shape resonance with non-resonant laser light}
\label{subsec:scheme}

\begin{figure}[tb]
  \centering
  \includegraphics[width=0.9\linewidth]{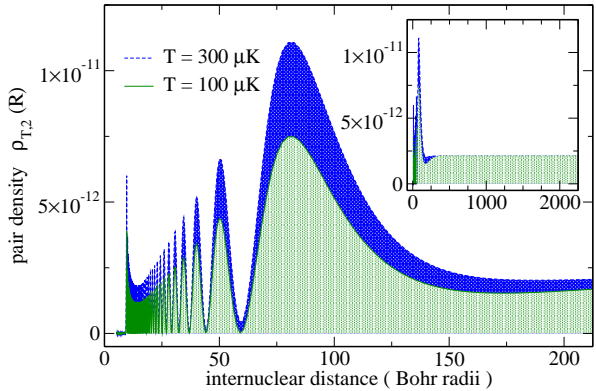}
  \caption{Thermal pair density $\rho_{T,2}(R)=\int
    d\cos\theta\,\rho_{T,2} (R,\theta)$ for $^{87}$Rb atoms (without
    any non-resonant field): at long
    range, the probability of finding two atoms at a certain distance
    is equally distributed; at short range the interaction potential
    leads to a modulation of the pair density. For $T=300\,\mu$K, the
    presence of the close-by $d$-wave shape resonance is reflected in an
    enhanced pair density at short range  compared to $T=100\,\mu$K
    and in the nodes of $\rho_{T,2}(R)$ disappearing, for example near
    $R=58\,$a$_0$, due to states with $J>0$. } 
  \label{fig:rho}
\end{figure}

Photoassociation in a magneto-optical trap can easily be saturated.
Its efficiency is limited by the pair density at or near
the Condon radius which is the internuclear distance where the 
photoassociation laser induces a resonant transition from the initial
pair of colliding atoms to a weakly bound level of the atoms' electronically
excited state. The $R$-dependence of the pair density is given by 
$\rho_{T,2}(R)=\int d\cos\theta\,\rho_{T,2} (R,\theta)$ and 
illustrated in Fig.~\ref{fig:rho} for $^{87}$Rb atoms. 
The constant behavior at large distances results from summing over
many plane waves with random phases and reflects the equiprobability of finding two atoms at a certain
distance. At short range, the interaction potential modulates the pair
density. If only $s$-waves contribute significantly to the pair
density, all scattering wave functions have their nodes 
at the same position at short range, giving rise to zeros in the pair
density, cf. $\rho_{T,2}(R)$ for $T=100\,\mu$K in Fig.~\ref{fig:rho}.
This reflects a threshold law~\cite{CrubellierJPB06}
governing collisions at very low energy.
Photoassociation occurs at intermediate distances, typically between
$40\,$a$_0$ and $200\,$a$_0$  depending on the excited state potential
curve. Such a choice of Condon radius is a compromise between large
free-to-bound transition probabilities (demanding large $R$) and 
sufficiently large binding energies required to avoid dissociation
back into two atoms upon return to the electronic ground state
(demanding short $R$). If formation of molecules in their electronic
ground state proceeds via a sequence of short, optimally chosen pump
and dump pulses, 
integrated pulse energies on the order of a few nano-Joule are
sufficient to completely deplete the pair density near the Condon
radius.~\cite{KochPRA06b} Nevertheless, this creates only  one
to ten molecules per pulse sequence, depending on the density of the atoms
in the trap.~\cite{KochJPhysB06} 

A further increase of the number of photoassociated molecules requires
manipulation of the initial pair density, $\rho_{T,2}(R)$.
To this end, we pursue the following idea for controlling a shape
resonance with non-resonant light: A shape resonance
is a quasi-bound state where the probability density of an atom pair
becomes  trapped inside the rotational barrier. For typical
detunings, this happens at distances that
correspond to the Condon radius for photoassociation~\cite{FrancoiseReview,JonesRMP06}  and can therefore enhance
the pair density that is available for excitation by a
photoassociation laser. However, due to the thermal weight,
$e^{-E_{nJ}/k_B T}$, a shape resonance
contributes significantly to the thermal ensemble described by
Eq.~\eqref{eq:rho_grid} only if its energy is close to the thermal
energy of the scattering atoms. However, the energy of shape
resonances is usually of the order of milli-Kelvin, i.e., one to
two orders of magnitude above typical MOT temperatures.
We therefore apply  non-resonant light with an intensity 
chosen such that the position of the shape resonance is moved toward
the mean energy of the scattering atoms. One option consists in
slowly switching on the non-resonant laser light, thereby 
ensuring  adiabatic following of the 
thermal cloud of atoms. The field-free eigenstates present in the
thermal density, Eq.~\eqref{eq:rho_grid}, are then transformed
adiabatically into field-dressed states, i.e., 
into the eigenstates of Hamiltonian~\eqref{eq:2D_Hamil}. 
This modified density matrix, made up of the non-resonant field-dressed
states, constitutes the initial state for photoassociation with short
laser pulses.~\cite{KochPRA06a,KochPRA06b,KochPRA08}
Assuming that an optimally chosen photoassociation pulse completely
depletes the modified pair density near the Condon radius -- as in the field-free case~\cite{KochPRA06b} -- the enhancement of the
number of  photoassociated molecules is simply given by the ratio of the field-dressed to the
field-free thermal pair densities. 

Short-pulse photoassociation under non-resonant field control involves
two timescales: (i) a slow timescale for the switching of the
non-resonant laser light, determined by the requirement of
adiabaticity with respect to the rotational motion, and (ii) a short
timescale for the sequence of pump and dump pulses, determined by the
requirement of a bandwidth that is optimal for photoassociation.
% There is no general bandwidth requirement, it only makes sense if
% optimal for what is specified
The short timescale corresponds to a few picoseconds for
transform-limited pulses (or a few tens of picoseconds for shaped
pulses). 
The slow timescale is determined by the rotational periods which are
found to be % $1.6\,$ns ($0.95\,$ns) and 
$82\,$ns  and $36.5\,$ns
for the rotational ground state %s within the first and 
of the last vibrational
band for rubidium  and strontium, respectively. 
While it is rather difficult to assign a rotational period to regular
scattering states, the quasi-bound character of a shape resonance implies
that its rotational period is well-defined. For the $d$-wave resonance
of rubidium and the $g$-wave resonance of strontium considered below,
we find rotational periods of $2\,\mu$s and $350\,$ns, respectively.
A crucial question to be answered below is whether the modified shape
resonances live long enough to allow for adiabatic switching of the
non-resonant laser light. 

\section{Results and discussion}
\label{sec:results}

We study the non-resonant field control of a shape resonance for $^{87}$Rb
and $^{88}$Sr at typical MOT temperatures, between $50\,\mu$K
and $150\,\mu$K for rubidium and about $20\,\mu$K for
strontium.~\cite{ZelevinskyPRL06} For colliding $^{87}$Rb atom pairs,
a $d$-wave shape resonance has been observed at about
$290\,\mu$K,\cite{BoestenPRA97} and for 
$^{88}$Sr, we find a $g$-wave resonance at about
$1.75\,$mK.  Below,
we investigate the influence of non-resonant laser light on the
position and lifetime of these two shape resonances and the resulting
enhancement of pair density at photoassociation distances. 
The feasibility of non-resonant field control of a shape resonance is
determined by (i) the intensity required to move the position of the
resonance close to energies corresponding to the MOT temperature and
(ii) the lifetime of the modified resonance, which must be sufficiently long
to enable adiabatic switching of the non-resonant field. 
The finite lifetime of the shape resonance is caused by
tunneling through the rotational barrier of the quasi-bound resonance
state. Following the intuitive picture suggested by
Fig.~\ref{fig:eff_pot}, where the energy of the field-dressed shape
resonance is plotted for two laser intensities,
the lifetime of the resonance is 
expected to increase with the non-resonant field intensity. 

Calculating resonance lifetimes at very low energies is a challenging
numerical problem.\cite{JoseReview} We have therefore employed two 
different methods to determine resonance lifetimes -- complex absorbing potentials~\cite{RissJPB93} and
the width of peaks of the rotational constants, $B_n=\langle n|
\frac{1}{2\mu\Op R^2}|n \rangle$, that occur for shape
resonances~\cite{LondonoPRA10}. Complex absorbing potentials are characterized by
two parameters, potential strength and width. Convergence with respect
to these two parameters can be verified by representing the
eigenvalues, $(E_n,-i\Gamma_n/2)$, 
in the complex plane.~\cite{RissJPB93} 
When the resonance occurs at very low energy, the potential
strength needs to be small and the width huge, implying a very
large spatial grid. Moreover, if the resonance is pushed below the
dissociation threshold, the lifetime becomes infinite. As this limit
is approached, it becomes increasingly difficult to achieve convergence of the
parameters for the complex absorbing potential. When using the second
method, the rotational constants, 
$B_n=\langle n| \frac{1}{2\mu\Op R^2}|n \rangle$, are plotted versus energy, $E_n$,
and fitted to a Lorentzian.~\cite{LondonoPRA10} As the resonance
position is shifted to lower energy by the non-resonant field, 
fewer box-discretized continuum states contribute to the peak such that
identification of the peak becomes increasingly difficult. 
For all field intensities, the results agree with each other at least
within 5\%.
This accuracy is sufficient for the order-of-magnitude estimate
that is needed to determine the feasibility of the control scheme. 

\subsection{Rubidium}
\label{subsec:Rb}

The potential energy curve for the lowest triplet state,
$a^3\Sigma_u^+$, of $^{87}$Rb$_2$, is obtained by smoothly connecting
\textit{ab initio} data at short-range~\cite{Park2001}  
with the asymptotic expansion $C_6/R^6+C_8/R^8+C_{10}/R^{10}$ at long
range where the $C_i$ coefficients are taken from
Ref.~\onlinecite{MartePRL02}. Polarizabilities based on \textit{ab
  initio} calculations are found in Ref.~\onlinecite{DeiglmayrJCP08}
and fit Silberstein's formula very well at distances larger than
$16\,$a$_0$.\footnote{For $R>16\,$a$_0$,  the relative error
  is below $1\%$ ($3\%$) for the perpendicular (parallel) component.} 
Not surprisingly, the short-range behavior of the polarizabilities does
not play any role in our study, i.e., calculations with the
polarizabilities from Ref.~\onlinecite{DeiglmayrJCP08} and
calculations with polarizabilities according to Silberstein's
expansion, Eq.~\eqref{eq:alpha}, with some cutoff value at short
distance give identical results. 

\begin{figure}[tb]
  \centering
  \includegraphics[width=0.9\linewidth]{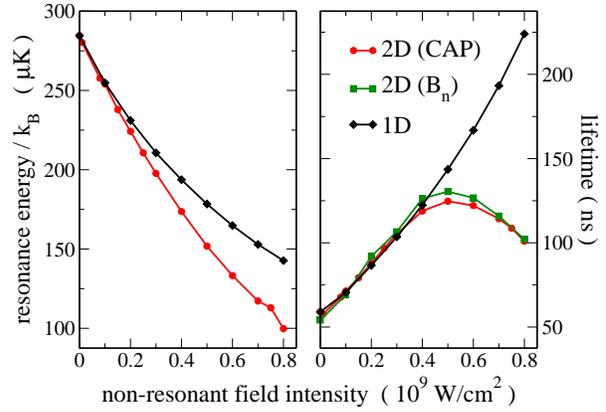}
  \caption{Energy and lifetime of the field-dressed 
    shape resonance vs intensity of the non-resonant field 
    for $^{87}$Rb (field-free $d$-wave
    shape resonance in the lowest triplet potential). Results for
    calculations based on the full $2D$ Hamiltonian, Eq.~\eqref{eq:2D_Hamil},
    and the adiabatic approximation, Eq.~\eqref{eq:1D_eff}, are
    compared. The lifetimes are calculated employing complex
    absorbing potentials (CAP, $1D$ and $2D$) and determining the peak
    width of the rotational constants ($B_n$).
  }
\label{fig:res_Rb}
\end{figure}
Resonance energy and lifetime as a function of the non-resonant field
intensity $I$ are shown in Fig.~\ref{fig:res_Rb}
for the $d$-wave shape resonance of $^{87}$Rb, comparing the results
of the $2D$ description obtained by diagonalizing the
Hamiltonian~\eqref{eq:2D_Hamil} with those obtained for the adiabatic
model, 
Eq.~\eqref{eq:1D_eff}. The position of the resonance is moved to
smaller energies by increasing the field intensity, as expected. 
Overall, only moderate field intensities, smaller than 
$10^9\,$W/cm$^2$, are required to move the position of the resonance
close to the energy that corresponds to a typical MOT temperature of
100$\,\mu$K. The
adiabatic approximation underestimates the energy shift significantly for
intensities larger than $3\times
10^8\,$W/cm$^2$. We attribute this disagreement between the adiabatic
approximation and the $2D$ description to the fact that states belonging to
different vibrational manifolds are mixed by a sufficiently
strong non-resonant field. Such a mixing is not accounted for in
the adiabatic approximation, cf. Section~\ref{subsec:model}.

While the adiabatic approximation, Eq.~\eqref{eq:1D_eff} captures 
the qualitative behavior of the resonance energy as a
function of non-resonant field intensity correctly, its disagreement with
the full $2D$ description, Eq.~\eqref{eq:2D_Hamil},  with respect to the
lifetimes is striking for the large intensities shown in 
Fig.~\ref{fig:res_Rb}. In fact, such a dramatic breakdown of the
adiabatic approximation is not necessarily expected since overall the
intensities of Fig.~\ref{fig:res_Rb} are moderate and one would expect
a quantitative disagreement such as that found for the
resonance energy. The adiabatic approximation essentially assumes that
the effect of hybridization due to the
non-resonant field is fully captured by the modified barrier height of
effective potential, cf. Fig.~\ref{fig:eff_pot}; and it neglects in
particular a coupling between rotational and vibrational motion. 
Vibrational motion that qualitatively differs for different rotational
states is thus not correctly accounted for. 

\begin{figure}[tb]
  \centering
  \includegraphics[width=0.9\linewidth]{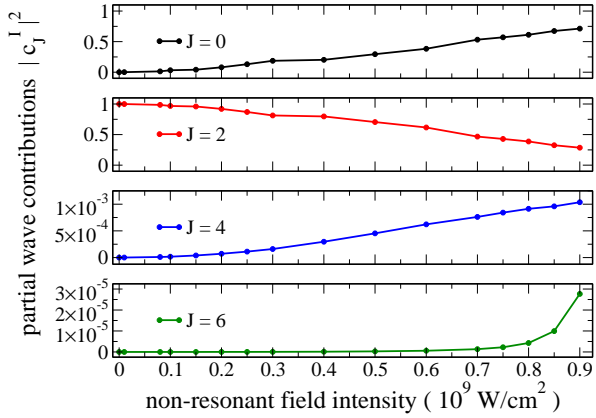}
  \caption{Contribution of different partial waves to the resonance
    wavefunction vs
    non-resonant field intensity for $^{87}$Rb 
    (field-free $J=2$ shape resonance in the lowest triplet potential).
  }
\label{fig:J_Rb}
\end{figure}
The degree of hybridization in the resonance wave function 
is examined in Fig.~\ref{fig:J_Rb} by plotting the 
absolute square of the rotational weights,
\[
c^I_J = \sum_n \int dR \int d \cos\theta \,\psi^{I\,*}_{res}(R,\theta)
\varphi_{nJ}(R,\theta)\,,
\] 
for the lowest four even partial waves as a function of non-resonant field
intensity (note that the coupling mixes only
partial waves of the same parity). Here, $\varphi_{nJ}(R,\theta)$ denote
the field-free eigenstates, i.e., the field-free box-quantized
scattering states, and $\psi^I_{res}(R,\theta)$ is the resonance
wavefunction obtained by diagonalizing
Hamiltonian~\eqref{eq:2D_Hamil} for a given value of the non-resonant
field intensity $I$. For $I=0$, the resonance is a pure $J=2$
state. As the non-resonant field intensity is increased, a substantial
amount of $J=0$ is mixed in. Higher partial waves do not contribute to
the resonance wavefunction (note the different scales of the $J=4$ and
$J=6$ panels). 
A turnover in the resonance lifetime as a function of field intensity is
observed in the  right-hand side of Fig.~\ref{fig:res_Rb}. At the
corresponding intensity, the $J=0$ contribution 
amounts to about 30\%, and for the largest
intensity shown in Fig.~\ref{fig:J_Rb}, the resonance wavefunction is
predominantly of $J=0$ character. The different behavior of the
lifetime in the full $2D$ Hamiltonian
and the adiabatic approximation is now
easily rationalized by the qualitatively different $R$-dependences of
the $J=2$ and $J=0$ states: While the field-free $J=2$ shape resonance
is a quasi-bound state confined to short range by the centrifugal
barrier, all $J=0$ states are of a purely scattering character.
As the non-resonant field intensity is switched on, the resonance
wavefunction becomes a superposition of these states, effectively
loosing its quasi-bound character. 

\begin{figure}[tb]
  \centering
  \includegraphics[width=0.9\linewidth]{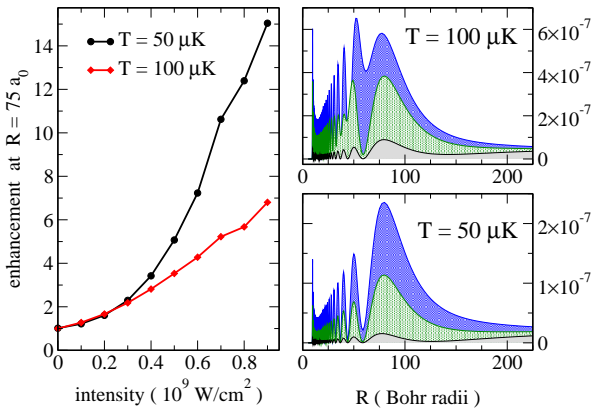}
  \caption{Modification of the rubidium pair density: Enhancement near
    $R=75\,$a$_0$ (left) and $R$-dependence of the field-dressed pair
    density, $\sum_{nJ}e^{-E_{nJ}/k_BT}
    \int d\cos\theta|\varphi_{nJ}(R,\theta)|^2/4\pi R^2$,
    cf. Eq.~\eqref{eq:rho_grid}, for zero field (grey), $I=6\times
    10^9\,$W/cm$^2$ (green) and  $I=9\times
    10^9\,$W/cm$^2$ (blue).
  }
\label{fig:enhance}
\end{figure}
The effect of the non-resonant field on the pair density is shown in
Fig.~\ref{fig:enhance}. The pair density at photoassociation distances
is enhanced by about one order of magnitude, cf. left-hand side of
Fig.~\ref{fig:enhance}, due to  
increasing the thermal weight of the shape resonance by moving its
energy close to $k_B$ times the trap temperature. 
Nevertheless, the enhancement is larger for $T=50\,\mu$K compared to
$T=100\,\mu$K although the resonance position is only moved to
$T=100\,\mu$K, cf. Fig.~\ref{fig:res_Rb} for the largest intensity
shown in Fig.~\ref{fig:enhance}. This is due to the fact that for 
$T=100\,\mu$K, already the field-free pair density is influenced by
the presence of the shape resonance, while its effect on the
field-free pair density is negligible for $T=50\,\mu$K. The change
brought about 
by the non-resonant field is thus larger for $T=50\,\mu$K. The enhancement of the pair
density at all short distances is illustrated by the right-hand side of
Fig.~\ref{fig:enhance} which shows the $R$-dependence of the
field-dressed pair density for three different intensities of the
non-resonant field.
The plotted quantity corresponds to the
  unnormalized pair density where the $\theta$-dependence has been
  integrated over. While a few partial waves are sufficient to
  converge the $R$-dependence of the pair density, the normalization
  factor requires a much larger number of partial waves for
  convergence. 
  
The pair density is significantly enhanced also at distances shorter
than the last peak of the last bound level. However, due to the fast
oscillations of the wavefunction, this pair density enhancement does
not translate into substantially larger number of molecules that can be
photoassociated. The largest increase of the photoassociation
efficiency is due to the pair density enhancement close to the last
peak of the last bound level, i.e., about $75\,$a$_0$ for rubidium. 

In summary, applying a non-resonant laser field shifts the position of
the $d$-wave rubidium shape resonance to smaller energies. While this
leads to an increasing lifetime of the resonance in an effective $1D$
model, the shape resonance looses its quasi-bound character in the
full $2D$ description since it is mixed with states of pure scattering
character. The lifetime therefore decreases after an initial
increase. The rubidium atom pair density at distances relevant to
photoassociation is found to increase by about one order of magnitude.

\subsection{Strontium}
\label{subsec:Sr}

The potential energy curve for the electronic ground state of
$^{88}$Sr$_2$ has been obtained spectroscopically, and an analytical
fit was reported in Ref.~\onlinecite{SteinPRA08}. We approximate the polarizabilities
by their long range expansion,
Eq.~\eqref{eq:alpha}, with the atomic values taken from
Ref.~\onlinecite{CRChandbook}. 

\begin{figure}[tb]
  \centering
  \includegraphics[width=0.9\linewidth]{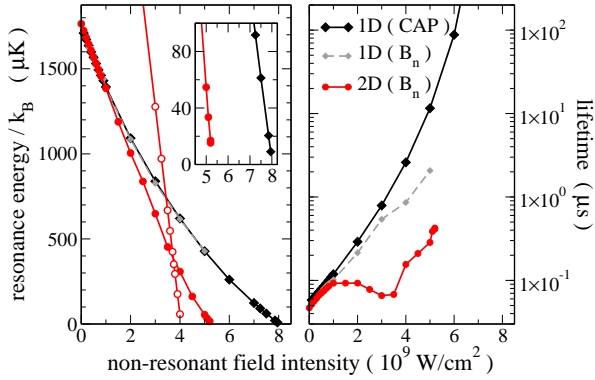}
  \caption{Energy and lifetime of the field-dressed 
    shape resonance vs intensity of the non-resonant field 
    for $^{88}$Sr (filled symbols: field-free $g$-wave ($J=4$)
    shape resonance in the electronic ground state potential, 
    empty symbols: field-free $J=8$ resonance). 
    Results for
    calculations based on the full $2D$ Hamiltonian, Eq.~\eqref{eq:2D_Hamil},
    and the adiabatic approximation, Eq.~\eqref{eq:1D_eff}, are
    compared. The lifetimes are calculated employing complex
    absorbing potentials (CAP, $1D$) and determining the peak
    width of the rotational constants ($B_n$, $1D$ and $2D$).
    Note the logarithmic scale for the lifetimes.
  }
\label{fig:res_Sr}
\end{figure}
Resonance energy and lifetime as a function of the nonresonant field
intensity $I$ are displayed in Fig.~\ref{fig:res_Sr}
for the $g$-wave shape resonance of $^{88}$Sr, comparing again results
obtained for the $2D$ Hamiltonian~\eqref{eq:2D_Hamil} with those
of the adiabatic approximation, 
Eq.~\eqref{eq:1D_eff}. Like in the case of rubidium,
cf. Fig.~\ref{fig:res_Rb}, the position of the resonance is moved to
smaller energies by increasing the non-resonant field intensity.
Slightly larger field intensities are required than for rubidium.
This is easily rationalized in terms of the different atomic
polarizabilities and rotational constants. The atomic polarizability
of rubidium is larger than that of strontium, yielding a stronger
interaction. Moreover, the rotational constant of strontium is larger 
than that of rubidium, so in order to achieve the same effect, a larger
non-resonant field is required. 
Nevertheless, the intensities remain moderate for strontium as well: about 
$5 \times 10^9\,$W/cm$^2$ are required to move the position of the
resonance close to the energy corresponding to a MOT
temperature of 20$\,\mu$K, typical for the two-color MOTs employed for 
alkaline-earth species.~\cite{ZelevinskyPRL06} 
Also, like for rubidium, 
the adiabatic approximation significantly underestimates the energy shift
at the large laser intensities. This becomes particularly
evident at intensities in excess of $3\times 10^9\,$W/cm$^2$.
Inspecting the pair density enhancement for strontium (see
Fig.~\ref{fig:enhanceSr} below), a second
resonant feature is observed which is attributed to a field-free $J=8$
shape resonance. The position of this resonance vs non-resonant field
intensity is traced by empty red circles
in Fig.~\ref{fig:res_Sr}. While the field-free $J=8$ resonance
starts out at an energy much larger than that of the $J=4$ resonance,
its energy is more strongly affected by the non-resonant field than
that of the $J=4$ one. It is therefore moved to lower temperatures
faster. The positions of the two resonances cross at about
$3.7 \times 10^9\,$W/cm$^2$. Since both states have the
same symmetry, the shape resonances exhibit an avoided crossing
as $I$ is varied. This
field-induced phenomenon is characterized by a strong coupling 
between the vibrational and rotational motion.\cite{gonzalez05_2} It
cannot be described by the
adiabatic model which treats these two degrees of freedom as independent.

While for both rubidium and strontium the field-free lifetimes are on
the order of 50$\,$ns, a dramatic increase of the lifetime is 
observed for strontium as the non-resonant field is applied (note the
logarithmic scale of the lifetimes on 
the right-hand ordinate in Fig.~\ref{fig:res_Sr}). This is in a stark
contrast with rubidium, cf. Fig.~\ref{fig:res_Rb}, where an increasing
lifetime is observed only in the adiabatic approximation,  but was
not confirmed by the $2D$ calculation. In strontium, the $2D$
lifetimes decrease with the non-resonant field intensity in an
intermediate intensity range, but as the intensity is further
increased, the lifetimes go up by an order of magnitude. 
The dip in the lifetimes, Fig.~\ref{fig:res_Sr}, occurs in a range
of non-resonant field intensities where the $J=4$ and $J=8$ resonances
are close in energy. 
Past the avoided crossing, the lifetime of the field-free $J=4$
resonance increases again
due to the mixing that takes place in the crossing region: The $J=8$
resonance is quite narrow and thus long-lived. Due to the
interaction of the two resonances, this character of the resonance is
partially transferred to the field-free $J=4$ resonance. 
Despite the fact that both the adiabatic
approximation and the $2D$ description predict an overall increase of the
lifetimes, the adiabatic
approximation ceases to be valid for strontium (like for rubidium) at intermediate
and large intensities since quantitatively both energies and lifetimes
do not agree with those of the $2D$ description: the more conservative
estimates of the lifetime obtained in the adiabatic approximation
by determining the peak width of the rotational
constants are still an order of magnitude larger than the $2D$
values. As for rubidium, the breakdown of the adiabatic approximation
is attributed to mixing of states from different vibrational
manifolds. 

\begin{figure}[tb]
  \centering
  \includegraphics[width=0.9\linewidth]{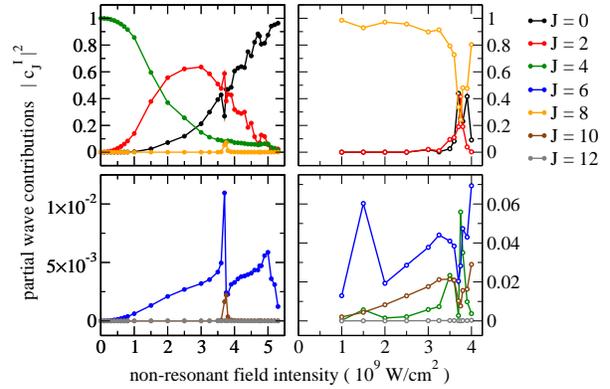}
  \caption{Contribution of different partial waves to the resonance
    wavefunction vs non-resonant field intensity for $^{88}$Sr 
    (field-free $J=4$ shape resonance on the left-hand side,
    field-free $J=8$ shape resonance on the right-hand side).
  }
\label{fig:J_Sr}
\end{figure}
A mixing of states from different vibrational
manifolds leads to particularly pronounced effects if two resonances are
concerned, i.e., for non-resonant field intensities close to
$I=3.7\times 10^9\,$W/cm$^2$, cf. left-hand side of Fig.~\ref{fig:res_Sr}. 
This is illustrated in Fig.~\ref{fig:J_Sr} in terms of the partial
 contributions to the 
resonance wavefunctions for both the
field-free $J=4$ resonance (left-hand side of Fig.~\ref{fig:J_Sr}) and 
the field-free $J=8$ shape resonance (right-hand side of
Fig.~\ref{fig:J_Sr}). 
For $I=0$, the resonance wavefunctions are pure
$J=4$ and $J=8$ states, respectively. 
As the non-resonant field is switched on, a substantial
amount of first $J=2$ and then $J=0$ is mixed in to the field-free
$J=4$ state. 
While this resonance wavefunction is predominantly of $J=2$ character
at intermediate intensities, it acquires essentially an $s$-wave
character at large intensities. 
Close to the intensity where the two resonances cross, 
some $J=8$ character is mixed in as well (little orange peak in the
left-hand side of Fig.~\ref{fig:J_Sr}). The presence of the second
resonance shows up in all the partial wave contributions causing
discontinuous behavior near the crossing point. 
Partial waves corresponding to $J=10$ and higher 
do not contribute to the resonance wavefunctions (grey symbols in the
left-hand side of Fig.~\ref{fig:J_Sr}). 
The field-free $J=8$ shape resonance (right-hand side of
Fig.~\ref{fig:J_Sr}) essentially keeps its character
with just a small admixture of $J=6$  until the crossing region with
the $J=4$ resonance is reached. There it acquires a substantial amount
of $J=2$ and $J=0$ character. 
This is due to the strong mixing between both states 
induced by the avoided crossing.
Once the crossing region is passed, the resonance wavefunction
recovers its mainly $J=8$ character.
This represents another unambigious signature of interaction between
the two resonances.  

\begin{figure}[tb]
  \centering
  \includegraphics[width=0.9\linewidth]{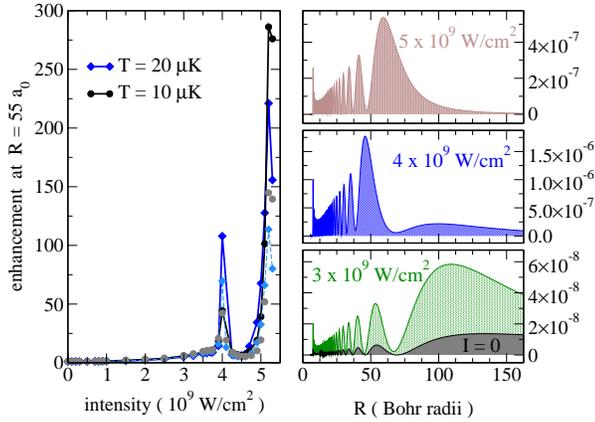}
  \caption{Modification of the strontium pair density: Enhancement
    near $R=55\,$a$_0$ (left) and 
    $R$-dependence of the field-dressed pair
    density, $\sum_{nJ}e^{-E_{nJ}/k_BT}
    \int d\cos\theta|\varphi_{nJ}(R,\theta)|^2/4\pi R^2$,
    cf. Eq.~\eqref{eq:rho_grid}, for $T=20\,\mu$K (right). 
    The lightblue and grey symbols in the left panel display the
    averaged enhancement (pair density  integrated from $R=32\,$a$_0$
    to $R=120\,$a$_0$.)    Note the
    different scales of the right-hand side panels.
  }
\label{fig:enhanceSr}
\end{figure}
The presence of the two resonances shows up clearly also in the
modification of the pair density, cf. Fig.~\ref{fig:enhanceSr}. 
While the first peak in the enhancement (left-hand side of
Fig.~\ref{fig:enhanceSr}), near $I=4 \times 10^9\,$W/cm$^2$, is due to the
field-free $J=8$
resonance whose energy approaches that corresponding to the trap
temperature, the second peak, near $I=5 \times 10^9\,$W/cm$^2$,
can be attributed to the field-free $J=4$ resonance.
Unlike rubidium, where the nodal structure of the pair density was
only slightly modified by shifting down the resonance,
cf. Fig.~\ref{fig:enhance}, the $R$-dependence of the pair density
depends sensitively on the non-resonant field intensity,
cf. right-hand side of Fig.~\ref{fig:enhanceSr}. The enhancement at a
specific position is therefore compared to an averaged enhancement
obtained by integrating the pair density in all relevant
photoassociation distances, from $R=32\,$a$_0$ to $R=120\,$a$_0$
(light-blue and grey symbols in the left-hand side of
Fig.~\ref{fig:enhanceSr}). The overall trend and order of magnitude
behavior of the averaged enhancement is the same as the enhancement at
a specific distance, $R=55\,$a$_0$. The sensitive $R$-dependence of
the pair density on laser intensity
reflects the contribution of the
two resonances which, in an effective description, correspond to two
different centrifugal barriers with different turning points. It is
the outer turning point at each of the centrifugal barriers that
causes the large peaks near $R=45\,$a$_0$ for $I=4 \times 10^9\,$W/cm$^2$ and
near   $R=60\,$a$_0$ for $I=5 \times 10^9\,$W/cm$^2$. 
Note that the pair density is also enhanced 
at the laser intensities where the two resonances exhibit the avoided crossing,
$I=3.75 \times 10^9\,$W/cm$^2$. However, the effect is most visible
for a temperature around $T=200\,\mu$K, which is much larger
than typical strontium MOT temperatures.

In summary, the interaction of two resonances leaves a pronounced
signature in the atom pair density, one  that should be observable in an
experiment which probes the pair density.~\cite{KochPRL09}
Moreover, for zero-field resonances which are far from
the trap temperature, a large enhancement of the atom pair density -- here
about 2 orders of magnitude -- and an increase in the field-free
resonance lifetime are found.

We note that an enhanced pair density at photoassociation internuclear
separations brought about by tuning a shape resonance could also
enable 
photoassociation based on Stimulated Raman Adiabatic Passage
(STIRAP)~\cite{BergmannRMP98} either within a single pair of pump and
Stokes pulses or within a sequence of phase-locked STIRAP 
pulse pairs.~\cite{ShapiroPRA07,ShapiroPRL08}
The feasibility of STIRAP photoassociation  depends 
on a sufficient isolation of the initial state from  
the scattering continuum. A possibility to achieve this 
discussed in the literature is based on utilizing  a
Feshbach resonance.~\cite{KuznetsovaNJP09} If, however, no resonance
is present, i.e., for an unstructured continuum, STIRAP fails.
Therefore, experiments that made use of STIRAP for transferring molecules
to their ground state started from molecules that had been already
associated via a Feshbach
resonance.~\cite{LangPRL08,OspelkausNatPhys08,NiSci08,DanzlSci08}     
A shape resonance that is brought to the right energy by non-resonant
laser light might provide an alternative way to isolate the  initial
state for STIRAP from the scattering continuum.
It would not rely on the presence of Feshbach resonances which do not
occur for example for the even isotope species of alkaline earth
atoms.~\cite{KoehlerRMP06}   
However, STIRAP induces a second slow timescale, due to the
requirement of adiabaticity with respect to the initial state. This
translates into an even longer lifetime of the shape resonance that is
required to guarantee success of STIRAP photoassociation
compared with the non-adiabatic picosecond pump-dump photoassociation
sequence discussed above.

\section{Conclusions}
\label{sec:concl}

We have considered the interaction of non-resonant laser light with
pairs of colliding atoms held in a magneto-optical trap, employing a
rigorous treatment of the thermal ensemble of atoms and comparing a
$2D$ description of the coupled rovibrational motion to an effective
$1D$ model based on the adiabatic approximation. 
Specifically, we have studied the influence of the non-resonant light
on the position and the lifetime of shape resonances. Such quasi-bound
states, that are trapped behind the centrifugal barrier imposed by the
angular momentum involved in the collision, lead to significant
enhancement of the pair density at comparatively short range. This 
enhancement readily translates into an increased efficiency of
molecule-formation processes such as
photoassociation.~\cite{BoestenPRL96,BoestenPRA97}  Since the 
photoassociation rate is limited by the pair density at or near the
Condon radius,~\cite{KochPRA06b,KochJPhysB06} 
utilizing shape resonances could overcome the main
obstacle toward forming large numbers of molecules. 

However, most of the time, the thermal weight of shape resonances at
MOT temperatures is very small because their energy is too high. In
the present study, we find 
that the energy of shape resonances can be decreased over
several orders of magnitude by applying a non-resonant laser field. It
is the second order nature of the laser-matter interaction that
guarantees a monotonous decrease in energy. Moderate intensities of
the non-resonant field are sufficient: intensities of less than
$10^9\,$W/cm$^2$ are required to move the $d$-wave resonance of
rubidium from $300\,\mu$K to below $100\,\mu$K, and intensities of
about $5\times 10^9\,$W/cm$^2$ shift the $g$-wave resonance of
strontium from 
$1.75\,$mK to below $20\,\mu$K. The value of
the required non-resonant field intensity is determined by the
rotational constant and the atomic polarizability of a given species:
small rotational constants and large polarizabilities are desirable to
maximize the matter-field interaction and minimize the (competing)
rotational kinetic energy. 

As the non-resonant field intensity is increased,
the lifetime of a shape resonance first increases, then
drops again. We rationalize the increase within the $1D$ picture by
the increase in the tunneling 
time through the centrifugal barrier which accompanies the lowering of the energy
of the resonance. However, the $1D$ picture captures only part of the
story: as the field intensity is further increased, strong
hybridization occurs. As scattering states belonging to different
partial waves get mixed in, the resonance wavefunction loses its
quasi-bound character,whereby tunneling out of the centrifugal barrier is
actually enhanced. This is the generic behavior of an isolated shape
resonance which is observed in our example of the $J=2$ resonance in
rubidium. A $2D$ description of the coupled rovibrational
motion is essential to capturing this effect correctly even for
comparatively moderate intensities. 
The adiabatic approximation, which
assumes  the vibrational and rotational degrees of freedom to be
independent for all states, ceases to be valid.

The physical situation changes
once more if more than one resonance comes into play, as in our
example of strontium where a broad $J=4$ resonance and a narrow 
$J=8$ resonance interact. For strong fields, the two resonances  show
an avoided crossing, in the course of which part of the
narrow-resonance character of the $J=8$ resonance is transferred to
the $J=4$ resonance, increasing the lifetime of the latter. 
Of course, the interaction of different resonances also requires a
$2D$ description of the coupled rovibrational motion and is not
accounted for in an effective $1D$ treatment.

For both single and interacting resonances, significant enhancement of the
thermal pair density at short distances is found. The magnitude
of the enhancement depends on the distance between the field-free
resonance position and the trap temperature. Maximum enhancement is
obtained when the non-resonant field intensity is chosen such that the
position of the shape resonance matches the trap temperature. In our
examples, an enhancement on the order of 10 and 100 was found for
rubidium and strontium, respectively. 
Such an enhancement can be probed directly by pulses that are short on the
timescale of the rotational and translational motion in the trap, for
example by picosecond pulses.~\cite{KochPRL09}

Our thermally averaged pair densities describe the structure of the
rovibrational eigenvalue problem for a given non-resonant field
intensity. This gives rise to both a dynamic and a static regime that 
can be pursued to implement non-resonant
field control of shape resonances in an experiment. The dynamic regime
consists of applying a  non-resonant laser pulse which is long
compared with the rotational motion associated with the shape resonance. The
shape resonance adiabatically follows the pulse and is shifted toward
lower energies. Once the energy corresponding to the trap temperature
is reached, a resonant photoassociation or probe pulse is applied that
is short with respect to the non-resonant laser pulse. This scheme
is expected to be successful if the timescale of the adiabatic switching
is of the order of the lifetime of the resonance or
shorter. In our study, we found rotational
periods of $2\,\mu$s and lifetimes of up to
$100\,$ns for rubidium, whereas for strontium we found  rotational periods of $350\,$ns and lifetimes of 500$\,$ns. Therefore, adiabatic following cannot be ensured for
rubidium. While losses due to the lifetime of the resonance might
% preclude is too strong
hamper the realization of a full two-orders-of-magnitude enhancement,
the enhancement found for strontium is significant. In general,
controlling a shape resonance with non-resonant laser light based on
adiabatic following will be successful for species with
short rotational periods (or large rotational constants). 

As an alternative to adiabatically shifting the position of a shape
resonance, a static
scheme can be pursued. This is based on the fact that the thermally
averaged atom pair density describes a thermal equilibrium in the trap for
a given temperature and non-resonant field intensity. The static
scheme involves application of a constant non-resonant field and a
hold period for thermaliziation of the atom cloud in the presence of
the field. The field-modified equilibrium can then serve as a
starting point for photoassociation, 
using either continuous-wave lasers or laser pulses. 

Finally, we wish to point out that our approach differs from previous
studies on the resonance-enhancement of the short-range pair density and the
subsequent molecule
formation~\cite{PellegriniPRL08,AlyabyshevPRA10,KremsPRL06,RosarioNJP09,ChakrabortyJPB11}  
in that we seek to \textit{actively} control the resonance. This is in
contrast to the literature which discusses the influence of a given resonance
on the desired process, e.g., molecule formation. 
Following the spirit of coherent control where resonances can be used
to imprint a well-defined phase onto the initial state of the desired
process,~\cite{ZemanPRL04} in our present study the resonance itself is
the subject of control, in order to afford an optimal use of it.
In terms of controllability, our main finding is that shape resonances
can be controlled in their energetic position but not at the same time
in their vibrational character (purely scattering vs
quasi-bound). This insight defines the challenge for future work
-- to identify a fully controllable scattering resonance.

\section*{Acknowledgments}

Financial support from the Deutsche Forschungsgemeinschaft
(Grant No. KO 2301/2), 
by the Spanish project FIS2008-02380 (MICINN) as well as the
Grants FQM-2445 and FQM-4643 (Junta de Andaluc\'{\i}a),
Campus de Excelencia Internacional Proyecto GENIL CEB09-0010  is
gratefully acknowledged. RGF belongs to the Andalusian research group
FQM-207.  BF and ML thank Gerard Meijer for discussions and support.

\footnotesize{
\bibliography{resonance} %your .bib file
\bibliographystyle{rsc} %the RSC's .bst file
}

\end{document}